\documentclass[twocolumn,english,aps,pre,showpacs]{revtex4}
\usepackage{amsmath}
\usepackage{amssymb}
\usepackage{graphicx}

\makeatletter
\usepackage{babel}
\makeatother

\begin{document}

\title{Effective surface shear viscosity of an incompressible particle-laden fluid interface}

\author{S.~V.~Lishchuk}

\affiliation{Department of Mathematics, University of Leicester, Leicester LE1 7RH, United Kingdom}


\begin{abstract}

The presence of even small amount of surfactant at the particle-laden fluid interface subjected to
shear makes surface flow incompressible if the shear rate is small enough
[T.~M. Fischer {\em et al}, J. Fluid Mech.\ {\bf558}, 451 (2006)].
In the present paper the effective surface shear viscosity of a flat, low-concentration,
particle-laden incompressible interface separating two immiscible fluids
is calculated. The resulting value is found to be 7.6\% larger than the value obtained
without account for surface incompressibility.

\end{abstract}

\maketitle


\section{\label{sec:introduction}Introduction}

To minimize total interfacial energy, suspended particles can self-assemble on the interface between
two fluids \cite{Binks:CPLI}. The structure and the dynamics of the particles at the interface are
controlled by their interaction, both direct ({\em eg} capillary, steric, electrostatic, magnetic,
van der Waals) and mediated by the surrounding fluids (hydrodynamic), as well as by size and
chemical composition of the adsorbed particles. Both the structure and the dynamics of the particles
contribute to their ability to stabilize particle-laden films, foams and emulsions, and to control
the transfer of matter through the interface \cite{Binks:2002:21,Vermant:2012:519,Mendoza:2013:0}.

At macroscopic scales, the particle-laden fluid interface can be viewed as a continuous infinitely
thin fluid interface with some {\em effective} static and dynamic properties, which can be measured
experimentally. Examples of static effective properties of particle-laden fluid interfaces are
surface tension \cite{Vignati:2003:6650}, bending \cite{Kralchevsky:2005:50} and saddle-splay
\cite{Lishchuk:2009:56001} elastic moduli. Dynamic properties include, in particular, surface
rheological parameters \cite{Mendoza:2013:0}.

The study of the surface rheology of particle-laden fluid interfaces can provide new insight into
their structure and properties. At the same time, rheology of these systems poses many experimental
and theoretical challenges. This makes it desirable constructing and investigating ``model
interfaces'', which could be utilized as reference systems in well defined limits, such as the
extremes of a purely Newtonian and purely elastic interface.

The rheology of a particle-laden interface becomes purely Newtonian when direct inter-particle
interactions are small, which is the case when adsorbed particles are separated widely enough. In
this case effective surface shear and dilatational viscosities arise due to viscous energy
dissipation in the surrounding bulk fluids. They can be calculated analytically for the special case
of the spherical particles having a solid-fluid contact angle (between particles' surface and
fluid-fluid interface) $\pi/2$, neglecting surface viscosity
\cite{Wilson:2002:425,Lishchuk:2009:016306}. The results can be used as a starting approximation for
more complicated systems in which the inter-particle interactions of a different nature cannot be
neglected.

The presence of adsorbed surfactant film on an interface introduces additional viscoelastic
stresses. If the concentration of surfactant is small, additional surface viscosity and elasticity
can be neglected. However, at small concentrations this case does not reduce to the case of
surfactant-free interface. The presence of even small amount of surfactant results in effective
surface incompressibility if the shear rate is small enough
\cite{Fischer:2006:451,Fischer:2004:139603}.

The effect of the incompressibility of the surface flow upon the effective surface shear viscosity
of a low-concentration particle-laden interface is calculated in this paper.


\section{\label{sec:model}Model}

We consider the steady flow of a system of identical rigid spherical particles of radius $R$
adsorbed at the flat interface between two incompressible fluids. We neglect gravity and assume the
interfacial tensions favor a contact angle $\pi/2$, so that the particles are located with the
equator coinciding with the interfacial plane.

We suppose the interface to be macroscopically thin, having surface tension high enough to
keep interface flat in the flow, and incompressible. These conditions correspond to small shear
rates which satisfy inequalities (\ref{eq:ra-tension}) and (\ref{eq:ra-incompressible}),
respectively.

We assume surface concentration of the adsorbed particles
\begin{equation}
\phi=\frac{\pi R^2N}A,
\end{equation}
with $N$ being the number of particle in area $A$, to be small, so that the motion of any particle
is not affected by other particles.

The energy dissipation rate in this system has two contributions, dissipation in the bulk fluids,
and the dissipation at the interface between two fluids.  We shall neglect surface dissipation. This
assumption can be expressed as Boussinesq number, defined as the ratio of surface viscosity of
the interface to the viscosity of bulk fluids,
\begin{equation}
\mathrm{Bo}=\frac{\eta_s}{(\eta_1+\eta_2)R},
\end{equation}
being small, $\mathrm{Bo}\ll 1$.

We assume both fluids 1 and 2 have the same densities, $\rho_1=\rho_2\equiv\rho$, and shear
viscosities, $\eta_1=\eta_2\equiv\eta$. We write the equations of motion in form valid for the
entire fluid domain as
\begin{equation}
\label{eq:equations-of-motion}
\boldsymbol\nabla\cdot\boldsymbol\sigma=0.
\end{equation}

The stress tensor $\boldsymbol\sigma$ has bulk and interfacial contributions,
\begin{equation}
\label{eq:stress-entire}
\boldsymbol\sigma=\boldsymbol\sigma_b+\delta(z)\boldsymbol\sigma_s.
\end{equation}
The bulk stress tensor,
\begin{equation}
\boldsymbol\sigma_b=-p\mathbf I
+\eta\left[\boldsymbol\nabla\mathbf v+(\boldsymbol\nabla\mathbf v)^\mathrm T\right]
\end{equation}
where $\mathbf v$ is velocity, $p$ is pressure, $\mathbf I$ is unit tensor, yields
Stokes equations \cite{Happel:LRNH},
\begin{equation}
\eta\nabla^2\mathbf v=\boldsymbol\nabla p,
\end{equation}
\begin{equation}
\label{eq:bulk-continuity}
\boldsymbol\nabla\cdot\mathbf v=0.
\end{equation}

In Eq.~(\ref{eq:stress-entire}) Dirac's delta function $\delta(z)$ is used to localize at the
plane $z=0$ the body force which arises from the jump conditions at the fluid interface
(see Ref.~\cite{Salac:2011:8192} and references cited therein).
We shall use the surface stress tensor in form
\begin{equation}
\boldsymbol\sigma_s=-\Pi\mathbf I_s,
\end{equation}
$\Pi$ being surface pressure, $\mathbf I_s=\mathbf I-\mathbf n\mathbf n$ being surface projector,
where we have neglected viscous contribution and dropped constant surface tension contribution since
it does not produce any body forces for the flat interface. The tangential stress balance at the
interface is
\begin{equation}
\mathbf I_s\cdot(\boldsymbol\sigma_2-\boldsymbol\sigma_1)\cdot\mathbf n
-\boldsymbol\nabla_s\Pi=0,
\end{equation}
where $\boldsymbol\nabla_s=\mathbf I_s\cdot\boldsymbol\nabla$ is surface gradient operator.

The above formulation corresponds to no-slip boundary condition at the fluid interface. We further
assume no-slip boundary condition at the surface of the particles. Surface continuity equation
for incompressible interfacial flow reads
\begin{equation}
\label{eq:surface-continuity}
\boldsymbol\nabla_s\cdot\mathbf v_s=0.
\end{equation}
Note that the component of the fluid velocity normal to the fixed interface should vanish,
\begin{equation}
\mathbf n\cdot\mathbf v_i=0.
\end{equation}

We shall consider the shear flow in the system with the motion of the fluid when unperturbed by
both particles and a fluid interface being described by the velocity field
\begin{equation}
\mathbf v^{(0)}=\boldsymbol\alpha\cdot\mathbf r,
\end{equation}
where the rate-of-strain tensor $\boldsymbol\alpha$ is symmetric, traceless, and constant. The
applied rate-of-strain is supposed to be small and the unperturbed pressure is taken to be zero.

It is well known that a shear flow distribution may be decomposed into a symmetric shear
and a rotation. By considering the motion in an appropriately chosen uniformly rotating rest frame,
we need concern ourselves solely with the dissipation associated with a symmetric shear. Taking into
account that the velocity at $z=0$ is parallel to the interfacial plane $XY$, and both bulk and
interfacial flow is incompressible, we can choose the symmetric rate-of-strain tensor to have the
form
\begin{equation}
\boldsymbol\alpha=
\begin{pmatrix}
\alpha & 0 & 0
\\
0 & -\alpha & 0
\\
0 & 0 & 0
\end{pmatrix}.
\end{equation}
The corresponding unperturbed velocity field is
\begin{subequations}
\label{eq:unperturbed-velocity}
\begin{align}
v^{(0)}_x&=\alpha x,
\\
v^{(0)}_y&=-\alpha y,
\\
v^{(0)}_z&=0.
\end{align}
\end{subequations}

The velocity, pressure, and surface pressure fields in presence of the particles and the fluid
interface will be respectively written as
\begin{align}
\mathbf v&=\mathbf v^{(0)}+\mathbf v^{(1)},
\\
p&=p^{(0)}+p^{(1)},
\\
\Pi&=\Pi^{(0)}+\Pi^{(1)}.
\end{align}
The stress tensor can be decomposed as
\begin{equation}
\boldsymbol\sigma=\boldsymbol\sigma^{(0)}+\boldsymbol\sigma^{(1)},
\end{equation}
where
\begin{equation}
\boldsymbol\sigma^{(0)}=-p^{(0)}\mathbf I
+\eta\left[\boldsymbol\nabla\mathbf v^{(0)}+\left(\boldsymbol\nabla\mathbf v^{(0)}\right)^\mathrm T\right]
\end{equation}
is the stress tensor of unperturbed flow, and
\begin{equation}
\boldsymbol\sigma^{(1)}=\boldsymbol\sigma_b^{(1)}+\delta(z)\boldsymbol\sigma_s
\end{equation}
with
\begin{equation}
\boldsymbol\sigma_b^{(1)}=-p^{(1)}\mathbf I
+\eta\left[\boldsymbol\nabla\mathbf v^{(1)}+\left(\boldsymbol\nabla\mathbf v^{(1)}\right)^\mathrm T\right],
\end{equation}
is the perturbation due to presence of particles and incompressible fluid interface.


\section{\label{sec:dissipation}Effective surface shear viscosity}

In order to find an expression for the effective surface shear viscosity of the particle-laden
interface, we shall compare expressions for the rate of viscous dissipation calculated in two ways.
First, we consider the system as homogeneous, having an effective continuum interface with effective
surface shear viscosity $\eta_s^\mathrm{eff}$. Second, we considering the flow in presence of
particles explicitly. Equating the energy dissipation rate in both cases will provide the expression
for $\eta_s^\mathrm{eff}$. In our derivation we shall closely follow the idea of Einstein for
calculating the effective shear viscosity of a three-dimensional dilute suspension of solid
spherical particles \cite{Einstein:1906:289,Einstein:1911:591} in a form presented by Batchelor
\cite{Batchelor:FD}.

Consider a sphere of radius $r_0$ and volume $V_0$. The area of the interface contained within this
volume is
\begin{equation}
A_s=\pi r_0^2.
\end{equation}

The rate at which forces do work on the external boundary $A_0$ of volume $V_0$ is
\begin{equation}
W=\int_{A_0}\left(\mathbf v^{(0)}\cdot\boldsymbol\sigma\cdot\mathbf n\right)dA.
\end{equation}
In the case of an effective, continuum interface with effective surface shear viscosity
$\eta_s^\mathrm{eff}$, this rate equals
\begin{equation}
W=W_0+4\eta_s^\mathrm{eff}\alpha^2A_s,
\end{equation}
where
\begin{equation}
W_0=\int_{A_0}\left(\mathbf v^{(0)}\cdot\boldsymbol\sigma^{(0)}\cdot\mathbf n\right)dA.
\end{equation}

We now proceed to determine the rate in the case when particles are explicitly accounted for.
We have:
\begin{equation}
W=W_0+\int_{A_0}\left(
\mathbf v^{(0)}\cdot\boldsymbol\sigma^{(1)}\cdot\mathbf n
\right)dA.
\end{equation}
The rates in both cases should be equal, therefore
\begin{equation}
\label{eq:equal-dissipation}
4\eta_s^\mathrm{eff}\alpha^2A_s=\int_{A_0}\left(
\mathbf v^{(0)}\cdot\boldsymbol\sigma^{(1)}\cdot\mathbf n
\right)dA.
\end{equation}

Following Batchelor \cite{Batchelor:FD}, we transform the surface integral over $A_0$ to the volume
integral over $V_0$:
\begin{align}
\nonumber
&\int_{A_0}\left(\mathbf v^{(0)}\cdot\boldsymbol\sigma^{(1)}\cdot\mathbf n\right)dA
\\
&=\int_{V_0}\left(
(\boldsymbol\alpha\cdot\mathbf r)\cdot(\boldsymbol\nabla\cdot\boldsymbol\sigma^{(1)})
+\boldsymbol\alpha:\boldsymbol\sigma^{(1)}
\right)dV
\\
\nonumber
&\quad-\int_{\sum A_p}\left[
(\boldsymbol\alpha\cdot\mathbf r)\cdot(\boldsymbol\sigma^{(1)}\cdot\mathbf n)
\right]dA,
\end{align}
where $\sum_{A_p}(\ldots)$ is sum of the surface areas of the particles contained within volume
$V_0$. Using the equations of motion, Eq.~(\ref{eq:equations-of-motion}), and the relation
\begin{align}
\int_{V_0}\boldsymbol\alpha:\boldsymbol\sigma^{(1)}dV
&=\int_{V_0}2\eta\left[
\boldsymbol\alpha:\left(\boldsymbol\nabla\cdot\mathbf v^{(1)}\right)
\right]dV
\\
\nonumber
&=-\int_{\sum A_p}2\eta\left(
\mathbf v^{(1)}\cdot\boldsymbol\alpha\cdot\mathbf n
\right)dA,
\end{align}
valid due to the boundary condition $\mathbf v^{(1)}=\mathbf 0$ at $A_0$, we obtain
\begin{align}
&\int_{A_0}\left(\mathbf v^{(0)}\cdot\boldsymbol\sigma^{(1)}\cdot\mathbf n\right)dA
={}
\\
\nonumber
&=\sum\int_{A_p}
\left[
(\boldsymbol\alpha\cdot\mathbf r)\cdot(\boldsymbol\sigma^{(1)}\cdot\mathbf n)
-2\eta(\boldsymbol\alpha:\mathbf v^{(1)}\mathbf n)
\right]dA,
\end{align}
which has the same form as the expression given by Batchelor \cite{Batchelor:FD}, except that the
stress tensor now contains interfacial part. The integral of the corresponding contribution over
the surface of a particle,
\begin{equation}
\int_{A_p}\left(\mathbf v^{(0)}\cdot\boldsymbol\sigma_s\cdot\mathbf n\right)\delta(z)dA,
\end{equation}
has form
\begin{equation}
-\alpha R^2\int_0^{2\pi}\Pi^{(1)}\cos2\varphi d\varphi.
\end{equation}

As a result, the equality of the viscous energy dissipation rates in two cases,
Eq.~(\ref{eq:equal-dissipation}), becomes
\begin{align}
&4\eta_s^\mathrm{eff}\alpha^2A_s=
-\alpha R^2N\int_0^{2\pi}\Pi^{(1)}(R)\cos2\varphi d\varphi+{}
\\
\nonumber
&{}+N\int_{A_p}\Big[
(\boldsymbol\alpha\cdot\mathbf r)\cdot(\boldsymbol\sigma_b^{(1)}\cdot\mathbf n)
{}-2\eta(\boldsymbol\alpha:\mathbf v^{(1)}\mathbf n)
\Big]dA,
\end{align}
yielding the following expression for the effective surface shear viscosity:
\begin{align}
\label{eq:etas-formula}
&\eta_s^\mathrm{eff}=
-\frac\phi{4\pi\alpha}\int_0^{2\pi}\Pi^{(1)}(R)\cos2\varphi d\varphi+{}
\\
\nonumber
&{}+\frac\phi{4\pi R^2\alpha^2}\int_{A_p}\Big[
(\boldsymbol\alpha\cdot\mathbf r)\cdot(\boldsymbol\sigma_b^{(1)}\cdot\mathbf n)
{}-2\eta(\boldsymbol\alpha:\mathbf v^{(1)}\mathbf n)
\Big]dA.
\end{align}
This expression contains hydrodynamic fields evaluated at the surface of a single particle,
half-immersed at an incompressible fluid interface and subjected to unperturbed shear flow given by
Eq.~(\ref{eq:unperturbed-velocity}).


\section{\label{sec:result}Result}

In order to obtain the value of the effective surface shear viscosity, we need to calculate the
integrals which enter Eq.~(\ref{eq:etas-formula}). For this, we need to know the hydrodynamic fields
at the surface of the particles. They can be obtained by solution of hydrodynamic equations for a
shear flow in the system with a single adsorbed particle.

The details of the numerical solution using vector spherical harmonics representation are given in
Appendix~\ref{sec:flow}. As a result of calculation, the effective surface shear viscosity is
\begin{equation}
\label{eq:K}
\eta_s^\mathrm{eff}=K\eta_{s0}^\mathrm{eff},
\end{equation}
where
\begin{equation}
\label{eq:halliday}
\eta_{s0}^\mathrm{eff}=\frac53(\eta_1+\eta_2)R\phi
\end{equation}
is the effective surface shear viscosity without account of the interface incompressibility
\cite{Lishchuk:2009:016306}, and the coefficient $K=1.076$.

In the rest of this section the conditions on the particle concentration and shear rate are given
under which the result should be valid. They follow from the model assumptions listed in
section~\ref{sec:model}.

Concentration of particles has to be small:
\begin{equation}
\label{eq:ra-dilute}
\phi\ll1.
\end{equation}
Lattice Boltzmann modeling of particle-laden flow without account of incompressibility has
demonstrated excellent agreement of the effective surface shear viscosity with
Eq.~(\ref{eq:halliday}) for the values of $\phi$ up to $\sim0.15$ \cite{Lishchuk:2009:016306}.

On the other hand, surface shear viscosity of the fluid interface can be neglected if it is small
compared to the effective surface shear viscosity due to particles. This gives another condition on
the concentration of particles:
\begin{equation}
\label{eq:ra-etas}
\phi\gg\frac{\eta_s}{\eta R}.
\end{equation}

Several model assumptions require shear rate to be small. First, Reynolds number has to be small to
allow neglecting inertial effects:
\begin{equation}
\label{eq:ra-reynolds}
\alpha\ll\frac\eta{\rho R^2}.
\end{equation}
Second, surface tension $\tau$ has to be large for the interface to be flat:
\begin{equation}
\label{eq:ra-tension}
\alpha\ll\frac\tau{\left|\eta_1-\eta_2\right|R}.
\end{equation}
Finally, surface compressibility $\kappa_s$ has to be small for the interface to be incompressible
\cite{Fischer:2006:451}:
\begin{equation}
\label{eq:ra-incompressible}
\alpha\ll\frac1{\kappa_s\eta R}.
\end{equation}


\section{\label{sec:conclusion}Concluding remarks}

The effective surface shear viscosity of a flat, low-concentration, particle-laden incompressible
interface separating two immiscible fluids has been found to be 7.6\% larger than the value obtained
without account for surface incompressibility.

Fischer {\em et al} \cite{Fischer:2006:451} calculated the drag on a sphere immersed and moving in
an incompressible monolayer. On a monolayer or membrane of low viscosity the translational drag on a
half-immersed sphere is 25\% larger than the drag on a sphere immersed in a free surface. The
present study shows that the effect of a low-viscosity monolayer on the effective viscosity of the
particle-laden fluid interface is less than on the drag coefficient of a sphere. Nevertheless, it is
not negligibly small and may be detected experimentally.

The present result was obtained in assumption of an infinite system size. The question remains open
whether other boundaries located at finite distance, which is the case for real systems, influences
the obtained value of the effective surface shear viscosity.

We finally note that the result for the surface dilatational viscosity of low-concentration
particle-laden interfaces \cite{Lishchuk:2009:016306} cannot be extended to the case of viscoelastic
fluid interface in the same way because the surface flow in this case is then essentially
compressible, which should be explicitly taken into account.

\acknowledgments

I thank Prof.\ Thomas Fischer for bringing this problem to my attention and fruitful discussions.


\appendix


\section{\label{sec:flow}Hydrodynamic fields}

In this appendix we solve hydrodynamic equations for a single particle at an incompressible
interface between two fluids in a shear flow given by Eq.~(\ref{eq:unperturbed-velocity}). The
general approach to the motion of particles in presence of an incompressible fluid interfaces was
developed by Blawzdziewicz and co-workers
Refs~\cite{Blawzdziewicz:1999:251,Blawzdziewicz:2010:114702,Blawzdziewicz:2010:114703}. To
facilitate the solution, we will make use of the symmetry of the problem and represent
hydrodynamic fields in vector spherical harmonics, and their two-dimensional analogue, vector polar
harmonics.


\subsection{Vector spherical harmonics}

Vector spherical harmonics are an extension of the scalar spherical harmonics for use with vector
fields. Vector spherical harmonics can be introduced in different ways
\cite{Hill:1954:211,Barrera:1985:287,Weinberg:1994:1086}. Vector spherical harmonics representation
is convenient for solving differential equations involving vector fields, such as equations of
hydrodynamics \cite{Friedman:2002:509} or electrodynamics
\cite{Barrera:1985:287,Carrascal:1991:184}.

We follow Ref.~\cite{Barrera:1985:287} and define vector spherical harmonics as
\begin{subequations}
\begin{align}
\mathbf Y_{lm}(\theta,\varphi)&=Y_{lm}\mathbf e_r,
\\
\boldsymbol\Psi_{lm}(\theta,\varphi)&=r\boldsymbol\nabla Y_{lm},
\\
\boldsymbol\Phi_{lm}(\theta,\varphi)&=\mathbf r\times\boldsymbol\nabla Y_{lm},
\end{align}
\end{subequations}
where $l=0,\ldots,\infty$, $m=-l,\ldots,l$, $\mathbf e_r$, $\mathbf e_\theta$ and
$\mathbf e_\varphi$ are orts in spherical coordinate system ($r$, $\theta$, $\varphi$), and scalar
spherical harmonics $Y_{lm}(\theta,\varphi)$ are defined as
\begin{align}
Y_{lm}(\theta,\varphi)&=
(-1)^m\sqrt{\frac{(2l+1}{4\pi}\frac{(l-m)!}{(l+m)!}}P_l^m(\cos\theta)e^{im\varphi},
\end{align}
where
\begin{equation}
P_l^m(x)=\frac{(-1)^m}{2^ll!}(1-x^2)^{m/2}\frac{d^{l+m}}{dx^{l+m}}(x^2-1)^l
\end{equation}
are associated Legendre polynomials \cite{Jackson:CE}.


\subsection{Vector polar harmonics}

To represent fields at a planar fluid interface we introduce vector polar harmonics in a way
analogous to vector spherical harmonics:
\begin{subequations}
\begin{align}
\mathbf y_m(\varphi)&=y_m(\varphi)\mathbf e_\rho,
\\
\boldsymbol\psi_m(\varphi)&=r\boldsymbol\nabla y_m(\varphi),
\end{align}
\end{subequations}
where $m=-\infty,\ldots+\infty$, and
\begin{equation}
y_m(\varphi)=\frac1{\sqrt{2\pi}}e^{im\varphi}
\end{equation}
are scalar polar harmonics.

Vector polar harmonics are related to vector spherical harmonics on a ${z=0}$ plane as
\begin{subequations}
\begin{align}
\mathbf Y_{lm}\left(\frac\pi2,\varphi\right)&=\chi_{lm}\mathbf y_m(\varphi)
\\
\boldsymbol\Psi_{lm}\left(\frac\pi2,\varphi\right)&=\chi_{lm}\boldsymbol\psi_m(\varphi)
\\
\boldsymbol\Phi_{lm}\left(\frac\pi2,\varphi\right)&=-\frac im\xi_{lm}\boldsymbol\psi_m(\varphi)
\end{align}
\end{subequations}
where the quantities $\chi_{lm}$ and $\xi_{lm}$ are defined by relations
\begin{equation}
Y_{lm}\left(\frac\pi2,\varphi\right)=\chi_{lm}y_m(\varphi)
\end{equation}
and
\begin{equation}
\left.\frac{\partial Y_{lm}(\theta,\varphi)}{\partial\theta}\right|_{\theta=\pi/2}=\xi_{lm}y_m(\varphi).
\end{equation}
The quantities $\chi_{lm}$ are explicitly given by formula
\begin{equation}
\chi_{lm}=\frac{(-1)^{(l-m)/2}}{2^l}\sqrt{\frac{2l+1}2}
\frac{\sqrt{(l+m)!(l-m)!}}{\left(\frac{l+m}2\right)!\left(\frac{l-m}2\right)!}
\end{equation}
for even $l+m$, and are zero otherwise. The quantities $\xi_{lm}$ are explicitly given by formula
\begin{align}
\xi_{lm}&=\frac{(-1)^{(l-m+1)/2}}{2^l}\sqrt{\frac{2l+1}2}
\nonumber\\
&\times(l+m+1)
\frac{\sqrt{(l+m)!(l-m)!}}{\left(\frac{l+m+1}2\right)!\left(\frac{l-m-1}2\right)!}
\end{align}
for odd $l+m$, and are zero otherwise. 


\subsection{\label{sec:representation}Representation of hydrodynamic fields}

The velocity field $\mathbf v^{(1)}(\mathbf r)$ is expanded in vector spherical harmonics as
\begin{align}
\mathbf v^{(1)}(\mathbf r)&=\sum_{lm}\left[
v^{(r)}_{lm}(r)\mathbf Y_{lm}(\theta,\varphi)
\right.
\nonumber\\
&\left.{}+v^{(1)}_{lm}(r)\boldsymbol\Psi_{lm}(\theta,\varphi)
+v^{(2)}_{lm}(r)\boldsymbol\Phi_{lm}(\theta,\varphi)
\right].
\end{align}

The pressure field $p^{(1)}(\mathbf r)$, being scalar, is expanded in scalar spherical harmonics:
\begin{equation}
p^{(1)}(\mathbf r)=\sum_{lm}p_{lm}(r)Y_{lm}(\theta,\varphi).
\end{equation}

The surface velocity field $\mathbf v_s(\mathbf r)$ is expanded in vector polar harmonics as
\begin{equation}
\mathbf v_s(\mathbf r)=\sum_{m=-\infty}^\infty\left[
u^{(\rho)}_m(r)\mathbf y_m(\varphi)+u^{(\varphi)}_m(r)\boldsymbol\psi_m(\varphi)
\right].
\end{equation}

The surface pressure field $\Pi^{(1)}(\mathbf r)$ is expanded in scalar polar harmonics:
\begin{equation}
\label{eq:surface-pressure-expansion}
\Pi^{(1)}(\mathbf r)=\sum_{m=-\infty}^{\infty}\Pi_m(r)y_m(\varphi).
\end{equation}


\subsection{Differential operators}

This subsection presents explicit expressions for the differential operators used in subsequent
derivation.

The expressions for the gradient of the scalar field, the divergence of the vector field, and the
curl of the vector field, expanded in vector spherical harmonics, are \cite{Barrera:1985:287}:
\begin{equation}
\boldsymbol\nabla p(\mathbf r)=\sum_{lm}\left[
\frac{\partial p_{lm}(r)}{\partial r}\mathbf Y_{lm}(\theta,\varphi)
+\frac{p_{lm}(r)}{r}\boldsymbol\Psi_{lm}(\theta,\varphi)
\right],
\end{equation}
\begin{align}
\label{eq:divergence}
&\boldsymbol\nabla\cdot\mathbf v^{(1)}(\mathbf r)=
\\
\nonumber
&\sum_{lm}\left[
\frac{\partial v^{(r)}_{lm}(r)}{\partial r}+\frac2rv^{(r)}_{lm}(r)-\frac{l(l+1)}rv^{(1)}_{lm}(r)
\right]Y_{lm}(\theta,\varphi),
\end{align}
\begin{align}
\boldsymbol\nabla&\times\mathbf v^{(1)}(\mathbf r)=\sum_{lm}\left\{
-\frac{l(l+1)}rv^{(2)}(r)\mathbf Y_{lm}(\theta,\varphi)
\right.
\\
\nonumber
&\left.{}-\left[\frac{\partial v^{(2)}(r)}{\partial r}+\frac1rv^{(2)}(r)\right]\boldsymbol\Psi_{lm}(\theta,\varphi)
\right.
\\
\nonumber
&\left.{}+
\left[
-\frac1rv^{(r)}_{lm}(r)+\frac{\partial v^{(1)}(r)}{\partial r}+\frac1rv^{(1)}_{lm}(r)
\right]\boldsymbol\Phi_{lm}(\theta,\varphi)\right\},
\end{align}
where
\begin{equation}
\sum_{lm}(\cdots)\equiv\sum_{l=0}^\infty\sum_{m=-l}^l(\cdots).
\end{equation}
Consequently, the Laplacian of a divergence-free vector field is
\begin{widetext}
\begin{align}
&\nabla^2\mathbf v^{(1)}(\mathbf r)=
-\boldsymbol\nabla\times\left[\boldsymbol\nabla\times\mathbf v^{(1)}(\mathbf r)\right]=
\sum_{lm}\left\{
-\frac{l(l+1)}r\left[
\frac{v^{(r)}_{lm}(r)}r
-\frac{\partial v^{(1)}_{lm}(r)}{\partial r}
-\frac{v^{(1)}_{lm}(r)}r
\right]\mathbf Y_{lm}(\theta,\varphi)
\right.
\\
\nonumber
&\quad{}-\left[
\frac1r\frac{\partial v^{(r)}_{lm}(r)}{\partial r}
-\frac{\partial^2v^{(1)}_{lm}(r)}{\partial r^2}
-\frac2r\frac{\partial v^{(1)}_{lm}(r)}{\partial r}
\right]\boldsymbol\Psi_{lm}(\theta,\varphi)
\left.
+
\left[
\frac{\partial^2v^{(2)}_{lm}(r)}{\partial r^2}
+\frac2r\frac{\partial v^{(2)}_{lm}(r)}{\partial r}
-\frac{l(l+1)}{r^2}v^{(2)}_{lm}(r)
\right]\boldsymbol\Phi_{lm}(\theta,\varphi)
\right\}.
\end{align}
\end{widetext}

The surface gradient of a scalar field and the surface divergence of a vector field are expanded in
polar harmonics as
\begin{align}
\label{eq:surface-gradient}
&\boldsymbol\nabla_s\Pi^{(1)}(\mathbf r)=
\\
\nonumber
&=\sum_{m=-\infty}^\infty\left[
\frac{\partial\Pi_m(r)}{\partial r}\mathbf y_{lm}(\varphi)
+\frac{\Pi_m(r)}{r}\boldsymbol\psi_{lm}(\varphi)
\right].
\end{align}
\begin{align}
\nonumber
&\boldsymbol\nabla_s\cdot\mathbf v_s^{(1)}(\mathbf r)=
\\
\label{eq:surface-divergence}
&=\sum_{m=-\infty}^\infty\left[
\frac{\partial v^{(\rho)}_m(r)}{\partial r}+\frac1rv^{(\rho)}_m(r)-\frac{m^2}rv^{(\varphi)}_m(r)
\right]y_m.
\end{align}


\subsection{Boundary conditions}

The boundary conditions at infinity $\mathbf v^{(1)}(\infty)=0$ and $p^{(1)}(\infty)=0$ can be
represented as
\begin{subequations}
\label{eq:boundary-conditions-infinity}
\begin{align}
v^{(r)}_{lm}(\infty)&=0,
\\
v^{(1)}_{lm}(\infty)&=0,
\\
v^{(2)}_{lm}(\infty)&=0,
\\
p_{lm}(\infty)&=0.
\end{align}
\end{subequations}

At particle surface we have $\mathbf v(R)=0$, which corresponds to
$\mathbf v^{(1)}(R)=-\mathbf v^{(0)}(R)$, or
\begin{subequations}
\begin{align}
v^{(1)}_x(R)&=-\alpha x,
\\
v^{(1)}_y(R)&=\alpha y,
\\
v^{(1)}_z(R)&=0.
\end{align}
\end{subequations}
These conditions are equivalent to
\begin{subequations}
\label{eq:boundary-conditions-particle}
\begin{align}
&u^{(r)}_{2,\pm2}(R)=-2\sqrt{\frac{2\pi}{15}}\alpha R,
\\
&u^{(1)}_{2,\pm2}(R)=-\sqrt{\frac{2\pi}{15}}\alpha R,
\\
&u^{(2)}_{2,\pm2}(R)=0,
\\
&u^{(r,1,2)}_{l,m}(R)=0 \quad \textrm{(for other values of $l$ and $m$).}
\end{align}
\end{subequations}


\subsection{Surface continuity equation}

By means of Eq.~(\ref{eq:surface-divergence}) for surface divergence of a velocity field, surface
incompressibility condition (\ref{eq:surface-continuity}) is equivalent to the relations
\begin{align}
\frac{\partial v^{(\rho)}_m(r)}{\partial r}+\frac1rv^{(\rho)}_m(r)-\frac{m^2}rv^{(\varphi)}_m(r)=0,
\\ \nonumber
m=-\infty,\cdots,+\infty.
\end{align}
Writing the surface velocity as
\begin{align}
\mathbf v_s(\mathbf r)&=\sum_{lm}\left\{
\chi_{lm}v^{(r)}_{lm}(r)\mathbf y_m(\varphi)
\right.
\nonumber\\
&\left.{}+\left[\chi_{lm}v^{(1)}_{lm}(r)-\frac im\xi_{lm}v^{(2)}_{lm}(r)\right]\boldsymbol\psi_m(\varphi)
\right\},
\end{align}
we can represent surface divergence of a velocity field as
\begin{align}
\boldsymbol\nabla_s\cdot\mathbf v_s(\mathbf r)=\sum_{lm}&\Bigg\{
\chi_{lm}\Bigg[\frac{\partial v^{(r)}_{lm}(r)}{\partial r}+\frac1rv^{(r)}_{lm}(r)
-{}
\\
\nonumber
&{}-\frac{m^2}rv^{(1)}_{lm}(r)\Bigg]
+\xi_{lm}\frac{im}rv^{(2)}_{lm}(r)
\Bigg\}y_m(\varphi).
\end{align}
The boundary conditions at the fluid interface thus become
\begin{align}
\nonumber
\sum_{l=|m|}^\infty\Bigg\{
\chi_{lm}\left[\frac{\partial v^{(r)}_{lm}(r)}{\partial r}+\frac1rv^{(r)}_{lm}(r)-\frac{m^2}rv^{(1)}_{lm}(r)\right]
+{}
\\
\label{eq:surface-contunuity-expansion}
{}+\xi_{lm}\frac{im}rv^{(2)}_{lm}(r)\Bigg\}
=0.
\end{align}


\subsection{Bulk continuity equation}

Since unperturbed flow satisfies the equation ${\boldsymbol\nabla\cdot\mathbf v^{(0)}(\mathbf r)=0}$,
bulk continuity equation (\ref{eq:bulk-continuity}) yields
\begin{equation}
\boldsymbol\nabla\cdot\mathbf v^{(1)}(\mathbf r)=0.
\end{equation}
Using Eq.~(\ref{eq:divergence}) for the divergence of the vector field expanded in vector spherical
harmonics, the continuity equation can be represented as
\begin{align}
\label{eq:bulk-contunuity-expansion}
\frac{\partial v^{(r)}_{lm}(r)}{\partial r}+\frac2rv^{(r)}_{lm}(r)&=\frac{l(l+1)}rv^{(1)}_{lm}(r),
\\
\nonumber
l=0,\cdots,\infty,
&\quad
m=-l,\cdots,l.
\end{align}


\subsection{Stokes equations}

The equations of motion (\ref{eq:equations-of-motion}) with the stress tensor given by
Eq.~(\ref{eq:stress-entire}),
\begin{equation}
\eta\nabla^2\mathbf v(\mathbf r)=
\boldsymbol\nabla p(\mathbf r)+\delta(z)\boldsymbol\nabla_s\Pi(\mathbf r),
\end{equation}
yield the equations for the perturbation fields only:
\begin{equation}
\eta\nabla^2\mathbf v^{(1)}(\mathbf r)=
\boldsymbol\nabla p^{(1)}(\mathbf r)+\delta(z)\boldsymbol\nabla_s\Pi^{(1)}(\mathbf r).
\end{equation}
We will write these equations in vector spherical harmonic representation.

We write the following expansions of the surface (Marangoni) force,
\begin{equation}
\label{eq:surface-force-definition}
\mathbf f(\mathbf r)=-\boldsymbol\nabla\Pi^{(1)}(\mathbf r) ,
\end{equation}
into vector (polar and spherical) harmonics:
\begin{equation}
\mathbf f(\mathbf r)=\sum_{m=-\infty}^\infty\left[
f^{(\rho)}_m(r)\mathbf y_m(\varphi)+f^{(\varphi)}_m(r)\boldsymbol\psi_m(\varphi)
\right],
\end{equation}
\begin{align}
\delta(z)\mathbf f(\mathbf r)&=\sum_{lm}\left[
f^{(r)}_{lm}(r)\mathbf Y_{lm}(\theta,\varphi)
\right.
\nonumber \\
&\left.{}+f^{(1)}_{lm}(r)\boldsymbol\Psi_{lm}(\theta,\varphi)
+f^{(2)}_{lm}(r)\boldsymbol\Phi_{lm}(\theta,\varphi)
\right].
\end{align}
Taking into account that
\begin{subequations}
\begin{align}
&\int_0^{2\pi}
\mathbf y_m(\varphi)\cdot\mathbf Y^*_{lm}\left(\frac\pi2,\varphi\right)
d\varphi=\chi_{lm},
\\
&\int_0^{2\pi}
\boldsymbol\psi_m(\varphi)\cdot\boldsymbol\Psi^*_{lm}\left(\frac\pi2,\varphi\right)
d\varphi=m^2\chi_{lm},
\\
&\int_0^{2\pi}
\boldsymbol\psi_m(\varphi)\cdot\boldsymbol\Phi^*_{lm}\left(\frac\pi2,\varphi\right)
d\varphi=im\xi_{lm},
\end{align}
\end{subequations}
we obtain the following relations:
\begin{subequations}
\begin{align}
f^{(r)}_{lm}(r)&=\frac{\chi_{lm}}rf^{(\rho)}_m(r),
\\
f^{(1)}_{lm}(r)&=\frac{m^2\chi_{lm}}{l(l+1)r}f^{(\varphi)}_m(r),
\\
f^{(2)}_{lm}(r)&=\frac{im\xi_{lm}}{l(l+1)r}f^{(\varphi)}_m(r).
\end{align}
\end{subequations}
Substituting Eq.~(\ref{eq:surface-pressure-expansion}) in Eq.~(\ref{eq:surface-force-definition}),
and using Eq.~(\ref{eq:surface-gradient}), we can write
\begin{subequations}
\begin{align}
f^{(r)}_{lm}(r)&=-\frac{\chi_{lm}}r\frac{\partial\Pi_m(r)}{\partial r},
\\
f^{(1)}_{lm}(r)&=-\frac{m^2\chi_{lm}}{l(l+1)r^2}\Pi_m(r),
\\
f^{(2)}_{lm}(r)&=-\frac{im\xi_{lm}}{l(l+1)r^2}\Pi_m(r).
\end{align}
\end{subequations}

Combining the above expressions, we obtain the following form of Stokes equations:
\begin{subequations}
\label{eq:stokes-expansion}
\begin{align}
\nonumber
-\eta\frac{l(l+1)}r\left[
\frac1rv^{(r)}_{lm}(r)
-\frac{\partial v^{(1)}_{lm}(r)}{\partial r}
-\frac1rv^{(1)}_{lm}(r)
\right]=
\\
=\frac{\partial p_{lm}(r)}{\partial r}
+\frac{\chi_{lm}}r\frac{\partial\Pi_m(r)}{\partial r},
\\
\nonumber
-\eta\left[
\frac1r\frac{\partial v^{(r)}_{lm}(r)}{\partial r}
-\frac{\partial^2v^{(1)}_{lm}(r)}{\partial r^2}
-\frac2r\frac{\partial v^{(1)}_{lm}(r)}{\partial r}
\right]=
\\
=\frac1rp_{lm}(r)
+\frac{m^2\chi_{lm}}{l(l+1)r^2}\Pi_m(r),
\\
\nonumber
\eta\left[
\frac{\partial^2v^{(2)}_{lm}(r)}{\partial r^2}
+\frac2r\frac{\partial v^{(2)}_{lm}(r)}{\partial r}
-\frac{l(l+1)}{r^2}v^{(2)}_{lm}(r)
\right]=
\\
=
\frac{im\xi_{lm}}{l(l+1)r^2}\Pi_m(r).
\end{align}
\end{subequations}


\subsection{\label{sec:numerical}Solution of hydrodynamic equations}

We are solving the Stokes equations, (\ref{eq:bulk-contunuity-expansion}) and
(\ref{eq:stokes-expansion}), and surface continuity equation,
(\ref{eq:surface-contunuity-expansion}), with boundary conditions
(\ref{eq:boundary-conditions-infinity}) and (\ref{eq:boundary-conditions-particle}).

By introducing functions $P_{lm}(r)$ satisfying $p_{lm}(r)=dP_{lm}(r)/dr$, we can write these
equations in form
\begin{subequations}
\label{eq:cauchy-euler}
\begin{align}
\frac{dv^{(r)}_{lm}(r)}{dr}+\frac2rv^{(r)}_{lm}(r)-\frac{l(l+1)}rv^{(1)}_{lm}(r)
=0,
\\
\eta\frac{l(l+1)}r\left[
\frac1rv^{(r)}_{lm}(r)
-\frac{dv^{(1)}_{lm}(r)}{dr}
-\frac1rv^{(1)}_{lm}(r)
\right]
\nonumber\\
{}+\frac{d^2P_{lm}(r)}{dr^2}
+\frac{\chi_{lm}}r\frac{d\Pi_m(r)}{dr}
=0,
\\
\eta\left[
\frac1r\frac{dv^{(r)}_{lm}(r)}{dr}
-\frac{d^2v^{(1)}_{lm}(r)}{dr^2}
-\frac2r\frac{dv^{(1)}_{lm}(r)}{dr}
\right]
\nonumber\\
{}+\frac1r\frac{dP_{lm}(r)}{dr}
+\frac{m^2\chi_{lm}}{l(l+1)r^2}\Pi_m(r)
=0,
\\
\eta\left[
\frac{d^2v^{(2)}_{lm}(r)}{dr^2}
+\frac2r\frac{dv^{(2)}_{lm}(r)}{dr}
-\frac{l(l+1)}{r^2}v^{(2)}_{lm}(r)
\right]
\nonumber\\
{}-\frac{im\xi_{lm}}{l(l+1)r^2}\Pi_m(r)
=0,
\\
\sum_{l=|m|}^\infty\left\{
\chi_{lm}\left[\frac{dv^{(r)}_{lm}(r)}{dr}+\frac1rv^{(r)}_{lm}(r)-\frac{m^2}rv^{(1)}_{lm}(r)\right]
\right.
\nonumber\\
\left.{}+\xi_{lm}\frac{im}rv^{(2)}_{lm}(r)
\right\}
=0,
\end{align}
\end{subequations}
where $l=2,\ldots,\infty$, and $m=\pm2$ due to boundary conditions at the surface of the particle
[Eq.~(\ref{eq:boundary-conditions-particle})]. This is an infinite system of Cauchy-Euler equations
for functions $v^{(r)}_{lm}(r)$, $v^{(1)}_{lm}(r)$, $v^{(2)}_{lm}(r)$, $P_{lm}(r)$ and $\Pi_m(r)$.
By means of substitution $r=e^t$ it is equivalent to an infinite system of linear homogeneous
differential equations with constant coefficients. The differential operator of this system can be
shown to be bounded \cite{Hille:1961:133}, therefore we can employ the method of reduction
\cite{Kantorovich:AMHA} and truncate the infinite system (\ref{eq:cauchy-euler}) at some value
$l=l_\mathrm{max}$. The solution of the infinite system will then be given by a limit of the
solution of the finite system at $l_\mathrm{max}\rightarrow\infty$.

As a result, the solution of the system (\ref{eq:cauchy-euler}) with boundary conditions
(\ref{eq:boundary-conditions-infinity}) and (\ref{eq:boundary-conditions-particle}) has form
\begin{subequations}
\begin{align}
&\Pi_m(r)=\sum_n\frac{c_{m,n}}{r^n},
\\
&v^{(r)}_{lm}(r)=\sum_n\frac{c^{(r)}_{l,m,n}}{r^n},
\\
&v^{(1)}_{lm}(r)=\sum_n\frac{c^{(1)}_{l,m,n}}{r^n},
\\
&v^{(2)}_{lm}(r)=\sum_n\frac{c^{(2)}_{l,m,n}}{r^n},
\\
&p_{lm}(r)=\sum_n\frac{c^{(p)}_{l,m,n}}{r^{n+1}},
\end{align}
\end{subequations}
where
\begin{equation}
n=2,3,5,7,9,\ldots,
\end{equation}
and the the values of the coefficients $c_{m,n}$, $c^{(r)}_{l,m,n}$, $c^{(1)}_{l,m,n}$,
$c^{(2)}_{l,m,n}$ and $c^{(p)}_{l,m,n}$ are provided by numerical calculation.

\begin{figure}[tb]
\begin{center}
\includegraphics[width=\columnwidth]{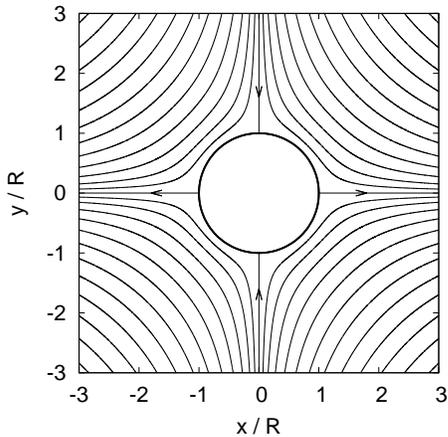}
\end{center}
\caption{\label{fig:flow}
Streamlines of the interfacial flow ($z=0$) near the particle.
}
\end{figure}

\begin{figure}[tb]
\begin{center}
\includegraphics[width=\columnwidth]{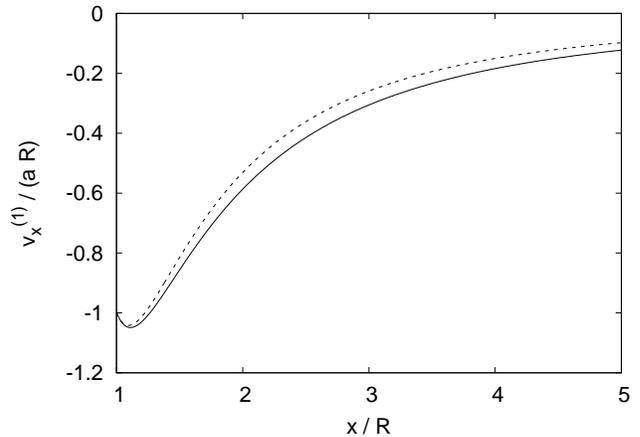}
\end{center}
\caption{\label{fig:velocity}
Perturbation velocity component $v^{(1)}_x$ at $y=0$, $z=0$ as a function of the distance $x$ from
the center of the particle with (solid line) and without (dashed line) account for incompressibility
of interfacial flow.
}
\end{figure}

This solution corresponds to a shear flow around a particle adsorbed at the incompressible fluid
interface (see Figure~\ref{fig:flow}). Velocity at large distance decays as $1/r^2$, but the
prefactor is greater due to incompressibility of the interfacial flow (see
Figure~\ref{fig:velocity}).


\subsection{Calculation of surface viscosity}

In order to obtain the value the effective surface shear viscosity, we need to calculate the
integrals which enter Eq.~(\ref{eq:etas-formula}). Substituting vector harmonic expansions of
hydrodynamic fields given in subsection~\ref{sec:representation}, we obtain
\begin{equation}
\label{eq:integral-surface}
\int_0^{2\pi}\Pi^{(1)}\cos2\varphi d\varphi=
\sqrt{2\pi}\Pi_2(R),
\end{equation}
\begin{align}
\nonumber
\int_{A_p}\Big[
(\boldsymbol\alpha\cdot\mathbf r)\cdot(\boldsymbol\sigma^{(1b)}\cdot\mathbf n)
{}-2\eta(\boldsymbol\alpha:\mathbf v^{(1)}\mathbf n)
\Big]dA={}
\\
\label{eq:integral-bulk}
{}=
\sqrt{\frac{32\pi}{15}}\bigg\{
v^{(r)}_{2,2}(R)+2\left[v^{(r)}_{2,2}\right]'(R)
-{}\quad
\\
\nonumber
{}
-9v^{(1)}_{2,2}(R)+3\left[v^{(1)}_{2,2}\right]'(R)
-p_{2,2}(R)
\bigg\},
\end{align}
where the prime indicates differentiation with respect to $r$, and the units are chosen such that
$\eta=1$, $\alpha=1$, $R=1$.

\begin{figure}[tb]
\begin{center}
\includegraphics[width=\columnwidth]{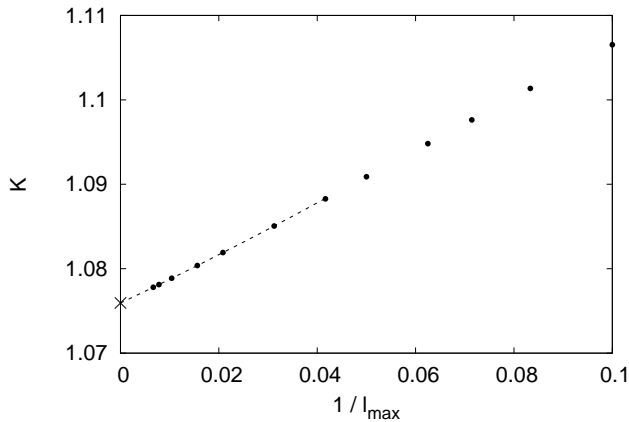}
\end{center}
\caption{\label{fig:convergence}
The coefficient $K$, defined by Eq.~(\ref{eq:K}), as a function of the truncation parameter
$l_\mathrm{max}$. Dashed line corresponds to quadratic extrapolation, cross indicates the value of
$K$ at $l_\mathrm{max}\rightarrow\infty$.
}
\end{figure}

Using the numerical solution described in subsection~\ref{sec:numerical}, we can calculate the
values of the expressions (\ref{eq:integral-surface}) and (\ref{eq:integral-bulk}) and,
consequently, the coefficient $K$, defined by Eq.~(\ref{eq:K}), as a function of the truncation
parameter $l_\mathrm{max}$ (see Figure~\ref{fig:convergence}). Numerical calculation of the limit
$l_\mathrm{max}\rightarrow\infty$ yields $K=1.076$.


\subsection{Viscosity contrast}

The above derivation used the assumption that the shear viscosities of both bulk fluids are equal.
In the case of different viscosities the hydrodynamic fields still satisfy the same equations and
boundary conditions. Therefore, the velocity field remains the same. Viscous dissipation in the
domain occupied by each bulk fluid will be proportional to the value of corresponding shear
viscosity, therefore the result given by Eq.~(\ref{eq:K}) remains valid in the case of velocity
contrast.





\end{document}